\documentclass[
  twocolumn,
  prl,
  showpacs,
  amsmath,
  amssymb,
  superscriptaddress,
  floatfix
]{revtex4}

\usepackage{bm}
\usepackage{dsfont}
\usepackage{graphicx}
\usepackage{pifont}
\usepackage{textcomp}
\usepackage{gensymb}
\usepackage{accents}
\usepackage{upgreek}

\renewcommand{\cite}[1]{{[}\onlinecite{#1}{]}}

\newcommand{\s}{\sum\limits}

\newcommand{\pa}{\partial}

\newcommand{\be}{\begin{equation}}
\newcommand{\e}{\end{equation}}
\newcommand{\beml}{\begin{subequations}}
\newcommand{\eml}{\end{subequations}}
\newcommand{\beq}{\begin{eqnarray}}
\newcommand{\eq}{\end{eqnarray}}
\newcommand{\ba}{\begin{array}}
\newcommand{\ea}{\end{array}}
\newcommand{\bpm}{\begin{pmatrix}}
\newcommand{\epm}{\end{pmatrix}}
\newcommand{\bc}{\begin{cases}}
\newcommand{\ec}{\end{cases}}
\newcommand{\lt}{\left}
\newcommand{\rt}{\right}
\newcommand{\n}{\nonumber}
\newcommand{\la}{\langle}
\newcommand{\ra}{\rangle}
\newcommand{\ep}{\varepsilon}

\newcommand{\bb}{\boldsymbol}

\DeclareMathOperator{\dv}{div}

\begin{document}

\title{Magnetoresistance in two-component systems}

\author{P.~S.~Alekseev}
\affiliation{A.\,F.\,Ioffe Physico-Technical Institute, 194021 St.\,Petersburg, Russia}
\author{A.~P.~Dmitriev}
\affiliation{A.\,F.\,Ioffe Physico-Technical Institute, 194021 St.\,Petersburg, Russia}
\author{I.~V.~Gornyi}
\affiliation{Institut f{\"u}r Nanotechnologie, Karlsruhe Institute of Technology, 76021 Karlsruhe, Germany}
\affiliation{A.\,F.\,Ioffe Physico-Technical Institute, 194021 St.\,Petersburg, Russia}
\author{V.~Yu.~Kachorovskii}
\affiliation{A.\,F.\,Ioffe Physico-Technical Institute, 194021 St.\,Petersburg, Russia}
\author{B.~N.~Narozhny}
\affiliation{\mbox{Institut f\"ur Theorie der Kondensierten Materie,
Karlsruhe Institute of Technology, 76128 Karlsruhe, Germany}}
\affiliation{National Research Nuclear University MEPhI (Moscow Engineering
Physics Institute), Kashirskoe shosse 31, 115409 Moscow, Russia}
\author{M.~Sch{\"u}tt}
\affiliation{School of Physics and Astronomy, University of Minnesota, Minneapolis, MN 55455, USA}
\author{M.~Titov}
\affiliation{
Radboud University Nijmegen, Institute for Molecules and Materials, NL-6525 AJ Nijmegen, The Netherlands}

\begin{abstract}
Two-component systems with equal concentrations of electrons and holes exhibit non-saturating, linear magnetoresistance in classically strong magnetic fields. The effect is predicted to occur in finite-size samples at charge neutrality in both disorder- and interaction-dominated regimes. The phenomenon originates in the excess quasiparticle density developing near the edges of the sample due to the compensated Hall effect. The size of the boundary region is of the order of the electron-hole recombination length that is inversely proportional to the magnetic field. In narrow samples and at strong enough magnetic fields, the boundary region dominates over the bulk leading to linear magnetoresistance. Our results are relevant for semimetals and narrow-band semiconductors including most of the topological insulators.
\end{abstract}


\maketitle

Growing interest in narrow-band semiconductors such as topological insulators and semimetals (e.g., graphene) continues to stimulate intense experimental research. An increasing number of these studies report observations of large linear magnetoresistance, which often shows no sign of saturation in classically strong magnetic fields even at room temperatures \cite{Friedman2010, Singh2012, Veldhorst2013, Wang2013, Gusev2013, Wiedmann2013}.

The story of linear magnetoresistance in non-magnetic compounds, notably in compensated semimetals \cite{Weiss1954}, can be traced back to the work by Kapitza in 1928 on the magnetoresistance of bismuth \cite{Kapitza1928}. The topic has received a revived attention after the discovery of huge linear magnetoresistance in bismuth films \cite{Yang1999,Yang2000} as well as in AgSe and AgTe compounds \cite{Xu1997, Husmann2002, Sun2003, Zhang2004, Hu2008}, which are narrow-band semiconductors \cite{Dalven1967}. A linear increase of resistance by three orders of magnitude has been seen in these experiments in a wide range of temperatures. The term ``titanic magnetoresistance'' has been coined very recently in Refs.~\cite{Liang2014, Ali2014, Wang2014, Pletikosic2014}, where both linear and non-linear change of resistance in CdAs, WTe, and NbSb has been observed.

Most of the conventional transport theories predict either absent or parabolic magnetoresistance. A theory of linear magnetoresistance in compensated Dirac semimetals has been proposed by Abrikosov back in 1969 \cite{Abrikosov1969}. His analysis is limited to the extreme quantum limit, $\omega_c \gg T$ (where $\omega_c$ is the cyclotron frequency, $T$ is the temperature, and $\hbar=k_B=1$). Still, linear magnetoresistance is routinely measured at room temperatures and in relatively weak magnetic fields for materials with very different spectra \cite{Friedman2010, Singh2012, Veldhorst2013, Wang2013, Gusev2013, Wiedmann2013, Weiss1954, Kapitza1928, Yang1999, Yang2000, Xu1997, Husmann2002, Sun2003, Zhang2004, Hu2008}. For these experiments the theory of Ref.~\cite{Abrikosov1969} does not apply and a purely classical explanation of the phenomenon has to be given.

One such explanation has been put forward by Parish and Littlewood on the basis of a classical random-resistor model \cite{Parish2003}, that was argued to describe a strongly inhomogeneous (or granular) material, such as AgSe. The model of Ref.~\cite{Parish2003}, however, makes no distinction between one- and two-component systems, while many of the aforementioned experimental studies stress that the presence of two types of charge carriers, e.g., electrons and holes, in nearly equal
concentrations is the necessary condition for the non-saturating, linear magnetoresistance to be observed \cite{Veldhorst2013, Zhang2004, Hu2008}. Furthermore, recent measurements of high-temperature linear magnetoresistance in high-quality BiSb nanosheets \cite{Veldhorst2013} and homogeneous monocrystalline HgTe/CdTe samples \cite{Gusev2013, Wiedmann2013} are inconsistent with the theory of Ref.~\cite{Parish2003}.

\begin{figure}
\centerline{
\includegraphics[width=\columnwidth]{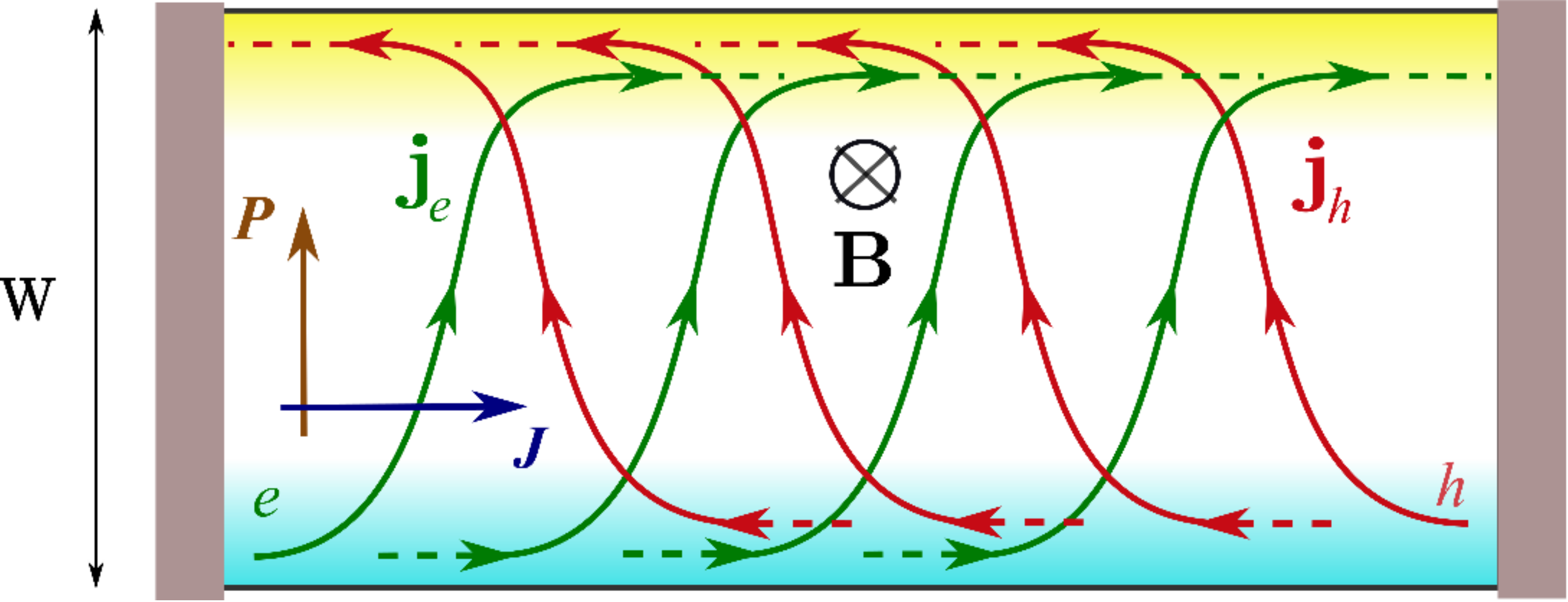}
}
\caption{Electron (green) and hole (red) trajectories in an electron-hole symmetric setup at charge neutrality. The bulk of the sample exhibits large geometric magnetoresistance as a consequence of the compensated Hall effect: electron and hole trajectories are tilted but the Hall voltage is absent. Lateral quasiparticle flow $\bb{P}$ results in excess quasiparticle density near the sample edges, where recombination processes due to electron-phonon interaction lead to linear magnetoresistance.}
\vspace{-0.5cm}
\label{fig:picture}
\end{figure}

In this Letter, we propose a classical mechanism for linear magnetoresistance in a two-component model. Our approach is based on the kinetic theory for a finite-size system near charge neutrality (charge compensation). The dominant contribution to the effect originates in the narrow regions near the sample edges, see Fig.~\ref{fig:picture}.

The conventional Drude theory \cite{Kittel1963} predicts that the longitudinal resistivity of a two-component system \cite{Weiss1954} depends on the applied magnetic field (in contrast to the simplest one-component case):
\be
\label{DrudeInf}
\rho_{xx}=\frac{\rho}{e\upmu}\frac{1+(\upmu B)^2}{\rho^2+n^2(\upmu B)^2}.
\e
Here $B$ is the magnetic field, $\upmu$ is the mobility (for simplicity, the mobility is taken to be the same for electrons and holes), $e$ is the absolute value of the electron charge, $\rho=n_e+n_h$ is the quasiparticle density, and $n=n_e-n_h$ is the charge density per unit charge with $n_{e(h)}$ standing for the corresponding electron (hole) densities. The equation~(\ref{DrudeInf}) predicts vanishing magnetoresistance far from the charge neutrality, $n=\rho=n_e$, and a non-saturating, quadratic magnetoresistance at charge neutrality, $n=0$, where the Hall effect is compensated: $\sigma_{xy}=\rho_{xy}=0$.

At charge neutrality, the above result corresponds to a constant quasiparticle flow, ${\bb{P}=\bb{j}_e+\bb{j}_h}$, which is orthogonal to the electric current $\bb{J}=-e\bb{j}=-e(\bb{j}_e-\bb{j}_h)$ (here $\bb{j}_e$ and $\bb{j}_h$ are the electron and hole current densities) due to the classical Hall effect. The lateral quasiparticle flow
$\bb{P}$ cannot be affected by the Hall voltage since the latter is not formed at charge neutrality. On the other hand, the quasiparticle current must vanish at the sample boundaries. Thus, the result of Eq.~(\ref{DrudeInf}) is strictly speaking incompatible with finite-size geometry.

Here we demonstrate that boundary effects may significantly modify Eq.~(\ref{DrudeInf}), leading to non-saturating, linear magnetoresistance near the charge neutrality point when the sample width is comparable with the electron-hole recombination length $\ell_0$. The latter may vary from hundreds of nanometers to centimeters depending on material properties and temperature, making the effect more important than previously anticipated. A similar phenomenon has been suggested to be responsible for a negative Coulomb drag in graphene at charge neutrality \cite{Titov2013}.

To develop intuition for the boundary effect let us consider a rectangular two-dimensional sample of the length $L$ and the width $W$, see Fig.~\ref{fig:picture}. For simplicity, we assume an electron-hole symmetric system at charge neutrality, where the electric current $\bb{J}$ is injected in $x$ direction.

Since the classical Hall effect for electrons compensates that for holes, the electrostatic potential in the sample remains flat and the charge density is zero everywhere, $n=0$. The distribution of electron and hole currents, $\bb{j}_{e,h}$, however, is non-trivial: it is essentially different in the bulk of the sample and in the boundary
regions, see Fig.~\ref{fig:picture}. In the bulk, the transversal quasiparticle current $\bb{P}=\bb{j}_e+\bb{j}_h$ leads to geometric magnetoresistance, $R_\textrm{bulk}=\frac{L}{W}\frac{1}{e\rho \upmu}(1+\upmu^2 B^2)$ [see Eq.~(\ref{DrudeInf}) for $n=0$]. In single-component systems such geometric effect is absent due to the presence of Hall voltage, unless the Corbino geometry is used or the sample is specifically prepared to be short and wide, i.e. for $W\gg L$ \cite{Weiss1954, Chang2014}.

\begin{figure}
\centerline{\includegraphics[width=0.9\columnwidth]{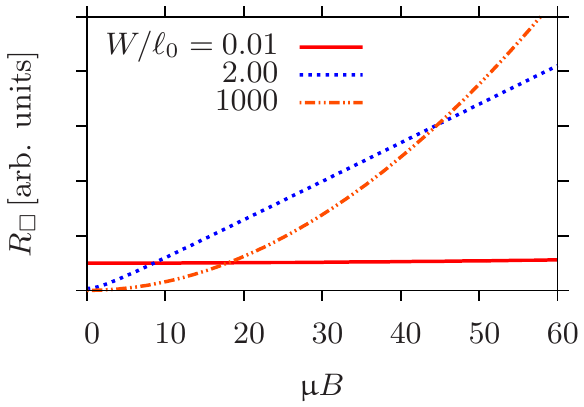}}
\caption{Sheet resistance $R_\square$ at charge neutrality versus magnetic field calculated from Eq.~(\ref{final0}) for three different values of the ratio $W/\ell_0$. 
The resistance is rescaled for a better presentation.  
}
\vspace{-0.5cm}
\label{fig:res0}
\end{figure}

The bulk current $\bb{P}$ leads to a formation of excess quasiparticle density near the sample edges at $y=\pm W/2$ (as shown in Fig.~\ref{fig:picture}) that has to be relaxed by electron-hole recombination, e.g., due to electron-phonon scattering.  This yields variation of the quasiparticle density over the distance $\ell_{R}=\ell_{0}/\sqrt{1+\upmu^2 B^2}$ from the boundary, where $\ell_{0}=2\sqrt{D\tau_{R}}$ depends on the diffusion coefficient $D$ and the recombination time $\tau_{R}$. With increasing magnetic field, the recombination length $\ell_{R}$ gets shorter because of multiple cyclotron returns of electron and holes to each other.

In the boundary regions of the size of the recombination length $\ell_{R}$ the electron and hole currents are directed essentially along the $x$-axis. Thus, the edge contribution to the overall resistance lacks the geometric enhancement and at charge neutrality is given by $R_\textrm{edge}=L/\ell_{R}e\upmu\rho$. The total sheet
resistance $R_\square$ of the sample with $W\gg\ell_{R}$  is estimated by regarding the edge and the bulk as parallel resistors: $R_\square^{-1}=(L/W)(R^{-1}_\textrm{bulk}+R^{-1}_\textrm{edge})$.  This yields
\be
\label{gen0}
R_\square=
\frac{1}{e\rho\upmu}
\lt(
\frac{1}{\upmu^2 B^2}+\frac{\ell_{0}}{W\,\upmu B}
\rt)^{-1},
\e
where we assumed $\upmu B\gg 1$ ($B=|\bb{B}|$). For sufficiently strong fields, the magnetoresistance at charge neutrality is linear in the field, namely
\be
\label{estimate}
R_\square=\frac{1}{e\rho}\frac{W}{\ell_{0}} B, \qquad
\frac{\ell_{0}}{\upmu B} \ll W\ll \upmu B \ell_{0}.
\e
Remarkably, within the semiclassical Drude picture any two-component neutral liquid is characterized by linear magnetoresistance as $B\to
\infty$.

Upon deviation from charge neutrality, the geometric resistance in the bulk of the sample disappears due to formation of the Hall voltage. From Eq.~(\ref{DrudeInf}) one finds
\be
\label{bulkdev}
R_\textrm{bulk}^{-1}=(W/L)\,e\rho\upmu\,\lt[1/(\upmu B)^2+n^2/\rho^2\rt],
\e
provided $\upmu B\gg 1$. Thus, the linear regime of Eq.~(\ref{estimate}) holds in strong fields as far as $n/\rho \ll \sqrt{\ell_0/W\upmu B}$.

In the remainder of this Letter we use the microscopic kinetic theory to show that our result~(\ref{estimate}) is generic for two-liquid systems at charge neutrality. We find that the effect can be realized both in disorder- and interaction- dominated regimes in materials with different spectra: in conventional narrow-band semiconductors with
parabolic spectrum (in particular, in the case when the symmetry between valence and conduction bands is violated) and in semi-metals with linear spectrum, e.g., in graphene. Technical details of the derivation are relegated to the Supplementary Material \cite{sup}.

Consider a model of a narrow-band semiconductor assuming for simplicity the parabolic spectra and energy-independent impurity scattering rates $\tau_{e,h}^{-1}$ for both electrons and holes,
\cite{sup}
\beml
\label{basic}
\beq
\label{n1}
D_e \bb{\nabla} \delta n_e + e \bb{E} n_{0,e}\tau_e /m_e - \bb{j}_e\times \bb{\omega}_e \tau_e =- \bb{j}_e, &&
\\
\label{n1h}
D_h \bb{\nabla} \delta n_h - e \bb{E} n_{0,h}\tau_h /m_h + \bb{j}_h\times \bb{\omega}_h \tau_h = -\bb{j}_h, &&
\\
\label{div1}
\dv \bb{j}_{e,h}=- (\Gamma_{e} \delta n_{e}+\Gamma_{h} \delta n_{h} ) /2, &&
\eq
\eml
where the index $\alpha=e,h$ refers to electrons or holes, $\bb{\omega}_{\alpha}={e\bb{B}}/{m_{\alpha} c}$, $D_{\alpha}$ is the averaged diffusion coefficient defined in \cite{sup}, $\delta n_{\alpha}(\bb{r})=n_{\alpha}(\bb{r})-n_{0,\alpha}$ is the density deviation from its equilibrium value $n_{0,\alpha}$, and $\Gamma_{\alpha}$ is the electron-hole recombination rate, e.g. due to electron-phonon interaction. For parabolic spectra, the cyclotron frequency $\omega_\alpha$ is independent of the chemical potential.

Equations (\ref{basic}) are justified most straightforwardly in the disorder-dominated regime, i.e. for ${\tau_\alpha\ll\tau_{\textrm{ee}}}$ (here $\tau_{\textrm{ee}}$ is
the inelastic electron-electron scattering time). The model~(\ref{basic}) ignores quantum effects: we assume $T\tau_\alpha \gg 1$, overlapping Landau levels $\omega_\alpha\ll T$, and, hence, the field-independent recombination rates $\Gamma_\alpha$. Similar equations can be derived for Dirac quasiparticles in graphene in the interaction-dominated regime \cite{long}. Consequently, the model~(\ref{basic}) is quite representative in a wide class of two-component systems.

In a narrow sample of length $L$ and width ${W\ll{L}}$, closed boundary conditions ${j_{y,\alpha}(y=\pm W/2)=0}$ lead to inhomogeneity of quasiparticle currents and densities. At charge neutrality, the electric charge remains uniform (due to the vanishing Hall effect). Away from the neutrality point, the charge density should be determined from a self-consistent solution of Eqs.~(\ref{basic}) and the corresponding electrostatic problem. In two-dimensional samples and in the limit of a strong screening by the gate electrode we may simplify the relation as
\be
\label{electro}
\bb{E}=E_0\bb{e}_{x}-\frac{e}{C}\frac{\pa \delta n}{\pa y}\bb{e}_{y},
\e
where $E_0$ is the external field, $\delta n=\delta n_{e}-\delta n_{h}$, $C=\epsilon/4\pi d$ is the gate-to-channel capacitance per unit area, $d$ is the the distance to the gate, $\epsilon $ is dielectric constant, and $\bb{e}_{x}$ is the unit vector in $x$ direction. In three-dimensional samples Eq.~(\ref{electro}) is replaced by $dE_y/dy=-e\,\delta n(y)/\epsilon d_0$, $E_x=E_0$, where $d_0$ is the sample thickness.

Further analysis is greatly simplified at the charge neutrality ($n_0=0$) under the assumption of electron-hole symmetry: $D_{\alpha}=D$, $m_{\alpha}=m$, $\Gamma_{\alpha}=1/\tau_R$, $\tau_{\alpha}=\tau$, ${\bb{\omega}_{\alpha}=\bb{\omega}_c=e\bb{B}/mc=\omega_c\bb{e}_z}$. 
We re-write Eqs.~(\ref{basic}) as
\beml
\label{basic2}
\beq
\label{n2a}
&&D\bb{\nabla}\delta \rho +\bb{P}- \bb{j} \times\bb{\omega}_{c}\tau=0,\\
\label{n3a}
&&D\bb{\nabla}\delta n +\bb{j}-e\bb{E} \rho_0\tau/m- \bb{P} \times\bb{\omega}_{c}\tau=0,\\
\label{div2b}
&&\dv \bb{P}=-\delta \rho/\tau_R, \qquad \dv\bb{j} =0,
\eq
\eml
where ${\delta\rho=\delta{n_{e}}+\delta{n_{h}}}$ is the deviation of the quasiparticle density from its equilibrium value ${\rho_0=n_{0,e}+n_{0,h}}$. The overall charge neutrality $n_0=n_{0,e}-n_{0,h}=0$ in the electron-hole symmetric system yields also the absence of charge fluctuations: $\delta n=0$. Thus, we find ${\bb{E}=E_0\bb{e}_x}$ irrespective of the electrostatic properties of the system.

The model of Eq.~(\ref{basic2}) is solved by $\bb{P}=P(y)\bb{e}_y$, $\bb{j}=j(y)\bb{e}_x$, $\delta\rho= \delta\rho(y)$. Excluding the variation of quasiparticle density from Eqs.~(\ref{basic2}), we rewrite the remaining equations as
\beml
\label{basic3}
\beq
\label{3a}
&&-D\tau_R\;\pa^2 P/\pa y^2 +P(y) + \omega_c \tau\,j(y) =0,\\
&&j(y)=j_0+\omega_c\tau\, P(y),
\label{3b}
\eq
\eml
where $j_0=e\tau\rho_0 E_0/m$ is the current in the absence of magnetic field. Excluding the current $j(y)$, we find the second-order differential equation for $P(y)$, which together with the boundary condition $P(\pm W/2)=0$ yields
\be
\label{profile}
P(y)=j_0\frac{\omega_c \tau}{1+(\omega_c\tau)^2}
\lt(\frac{\cosh(2y/\ell_R)}{\cosh(W/\ell_R)}-1\rt).
\e
Here we introduced the electron-hole recombination length $\ell_R=2\sqrt{D\tau_R/[1+(\omega_c\tau)^2]}$.

The result (\ref{profile}) and the corresponding current $j(y)$ obtained from Eq.~(\ref{3b}) are in full agreement with the qualitative distribution of quasiparticle currents shown in Fig.~\ref{fig:picture}. For a 2D sample the sheet resistance is defined as 
\be
\label{Rsquare}
R_\square=E_0/\overline{J},\qquad \overline{J}=-\frac{e}{W}\int\limits_{-W/2}^{W/2} j(y)\,dy.
\e
From Eqs.~(\ref{3b}), (\ref{profile}), and (\ref{Rsquare}) we obtain
\be
\label{final0}
R_\square=\frac{m}{e^2\tau\rho_0}\frac{1+(\omega_c\tau)^2}{1+(\omega_c\tau)^2 F(W/\ell_R)},
\e
where $F(x)={\tanh(x)}/{x}$. The result of Eq.~(\ref{final0}) is plotted schematically in Fig.~\ref{fig:res0} for three different values of the ratio $W/\ell_0$, where $\ell_0=2\sqrt{D\tau_R}$.

Let us analyse Eq.~(\ref{final0}) in three different regimes determined by the ratio of the sample width and the recombination length. For the widest samples, we find non-saturating geometric magnetoresistance, which is quadratic in the field \cite{Weiss1954}
\be
R_\square=\frac{m}{e^2\tau\rho_0}\lt[1+(\omega_c \tau)^2\rt], \qquad W\gg (\omega_c \tau)^2\ell_{R},
\e
where the geometric enhancement is the direct consequence of the compensated Hall effect: the electron and hole trajectories are tilted, but the Hall voltage is absent.

For the most narrow samples the geometric factor is absent
\be
R_\square=\frac{m}{e^2\tau\rho_0}, \qquad W\ll \ell_{R}.
\e
In this case, both electron and hole currents flow along the $x$-axis due to strong electron-hole recombination.

In classically strong magnetic fields, $\omega_c\tau\gg 1$, there exists another regime of intermediate system widths, where resistance depends linearly on the magnetic field:
\be
\label{linear1}
R_\square=\frac{m}{e^2\tau\rho_0} \frac{W}{\ell_{R}} \propto B,
\quad\ell_{R} \ll W\ll  (\omega_c \tau)^2\ell_{R}.
\e
This result is identical to Eq.~(\ref{estimate}) (note that $\omega_c\tau= \upmu B$).

The solution of Eqs.~(\ref{basic}) in the absence of the electron-hole symmetry and away from charge neutrality is more cumbersome (see Ref.~\cite{sup}), but the principle qualitative conclusions remain the same: close to charge neutrality the system exhibits linear magnetoresistance provided $\ell_0/\upmu B \ll W \ll \ell_0 \upmu B$, where $\upmu=\upmu_e\upmu_h/(\upmu_e+\upmu_h)$ is the average mobility of electrons and holes. Note that charge neutrality in the non-symmetric case no longer corresponds to the vanishing Hall resistance. For small deviations from neutrality point Eq.~(\ref{linear1}) becomes
\be
\label{linear2}
R_\square=\frac{m}{e^2\tau\rho_0} \frac{1}{\ell_{R}/W+\xi},
\quad\ell_{R} \ll W\ll  (\omega_c \tau)^2\ell_{R},
\e
where $\xi={n_0^2}/{\rho_0^2}$. The result (\ref{linear2}) can also be obtained from Eqs.~(\ref{gen0}) and (\ref{bulkdev}). Thus, the magnetoresistance is strongly peaked at charge neutrality.

The above results can be easily generalized to three-dimensional samples, arbitrary spectra, and the interaction-dominated regime \cite{long,sup}. The main conclusion remains robust: at the neutrality point, the system shows linear magnetoresistance in not too small magnetic field $B$.

\begin{figure}
\centerline{\includegraphics[width=0.9\columnwidth]{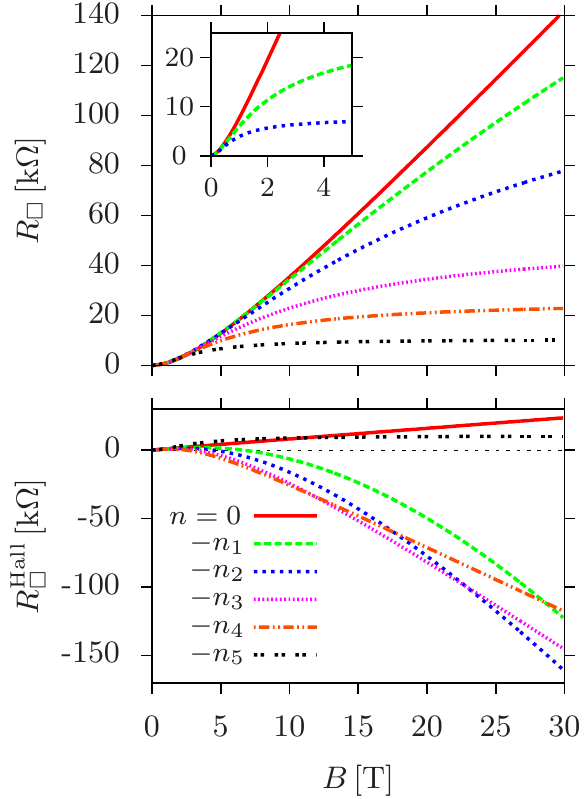}}
\caption{Sheet resistances $R_{\square}$ (top) and $R_{\square}^{\textrm{Hall}}$ (bottom) versus magnetic field for a two-dimensional narrow-band semiconductor with broken electron-hole symmetry: $\upmu_e=20\,\upmu_h$,  $\upmu_h=1$\,m$^2/$Vs at charge neutrality for $T=300$\,K. The solid line corresponds to $n=0$, while the other lines correspond to different negative densities $n=-n_i$ with $n_i=0.3, 0.5, 0.9, 1.3, 2.1\,\times$10$^{11}$\,cm$^{-2}$ for $W=10$\,$\mu$m, $d=0.5$\,$\mu$m, and $\epsilon_r=5$. The inset at the top panel shows the magnetoresistance $R_\square$ for a symmetric model with $\upmu_\alpha=20$\,m$^2/$Vs.
}
\label{fig:results}
\end{figure}

In Fig.~\ref{fig:results} we illustrate the magnetic field dependence of $R_\square$ and $R^{\textrm{Hall}}_{\square} = \overline{E_y}/\overline{J}$ obtained from the solution of Eqs.~(\ref{basic}) for some realistic parameters which correspond to a two-dimensional narrow-band semiconductor without electron-hole symmetry.  We consider a generic two-band model with the energy gap $E_g=4$\,meV at room temperature $T=300$\,K assuming different mobilities and velocities of electrons and holes at charge neutrality: $\upmu_e=20\,\upmu_h$, $\upmu_h=1$\,m$^2/$Vs, $v_e=10^6$\,m$/$s, $v_h=0.5\, v_e$. The plots correspond to different values of doping for a sample with $W=10$\,$\mu$m, $d=0.5$\,$\mu$m, and $\epsilon_r=5$. The recombination length in the absence of magnetic field equals $\ell_0=0.37$\,$\mu$m at charge neutrality.

In conclusion, we proposed a classical, recombination-induced mechanism of magnetoresistance in compensated semimetals and narrow-band semiconductors. The universal linear-in-$B$ magnetoresistance arises in finite-size samples in classically strong magnetic fields due to the interplay of bulk and edge contributions. This mechanism is expected to be relevant for explanation of linear magnetoresistance observed experimentally in various two-component systems. Our theory can be further extended to inhomogeneous samples in a spirit of Ref.~\cite{Knap2014}. One may expect that the linear magnetoresistance will take place in infinitely large systems
at charge neutrality provided the typical size of inhomogeneities is of the order of the zero-field recombination length $\ell_0$. Another possible generalization of our theory involves excitonic correlations between electrons and holes. These refinements will be presented elsewhere.

We are grateful to Sergey Roumyantsev, Yuri Vasil'ev, Steffen Wiedmann and Uli Zeitler for helpful discussions. The work was supported by the Dutch Science Foundation NWO/FOM 13PR3118, the EU Network Grant InterNoM, DFG-SPP 1459 and DFG-SPP 1666, GIF, the Humboldt Foundation, the Russian Foundation of Basic Research and the Dynasty Foundation.


\onecolumngrid
\setcounter{enumiv}{0} 
\setcounter{equation}{0} 
\setcounter{section}{0} 

\vspace{1cm}
\centerline{\bfseries ONLINE SUPPORTING INFORMATION}

\begin{quote}
In the supporting information we provide the technical details which are missing in the text of the Letter. The kinetic equation is formulated and applied to justify the linear response balance equations appeared in the main text. A general solution to the balance equations is also given.\\ 
\end{quote}

\maketitle

\maketitle

\twocolumngrid

\section{Kinetic equation}

Let us start with a stationary kinetic equation for a two-component liquid in two dimensions
\be
\label{A4}
{\bb{v}}_\alpha \frac{\pa f_\alpha}{\pa \bb{r}} + e_\alpha \lt(\bb{E}+\bb{v}_\alpha \times \bb{B} \rt) \frac{\pa f_\alpha}{\pa \bb{p}} = {\rm St}[f_\alpha],
\e
where $\alpha=(e,h)$ numerates the electron and hole components, $\bb{v}_\alpha =\pa \ep_\alpha(\bb{p})/\pa \bb{p}$, $e_h=-e_e=e$, $\bb{E}$ and $\bb{B}$ are the electric and magnetic fields, and $f_\alpha = f_\alpha(\ep,\hat{\bb{p}},\bb{r})$ is the distribution function which may depend on energy, momentum direction and coordinate. We use the units with $\hbar=c=1$, where $c$ is the speed of light. 

The collision integral at the right hand side of Eq.~\eqref{A4} describes the scattering by impurities and phonons as well as the electron-electron scattering. We consider two physically different situations: the case when impurity scattering dominates over the electron-electron and electron-phonon scattering, and the opposite case of very fast electron-electron collisions such that hydrodynamic approach can be applied. We aim to demonstrate that in both limits the two-component system shows linear-in-$B$ magnetoresistance at charge neutrality for sufficiently strong magnetic field $B$, even though the details of intermediate calculations appear to be quite different.

\section{Impurity-dominated regime}

\subsection{Exact \textit{e-h} symmetry and charge neutrality}

In this Section we assume that impurity scattering dominates over the electron-phonon and electron-electron scattering, i.e.
\be
\tau_{\textrm{imp}}^{-1} \gg \tau^{-1}_{\textrm{e-e}},\; \tau_{\textrm{ph}}^{-1},
\label{A11}
\e
where $\tau_{\textrm{imp}}$, $\tau_{\textrm{e-e}}$, and $\tau_{\textrm{ph}}$ are the impurity, electron-electron, and electron-phonon scattering times, correspondingly.
In this case the momentum relaxation is fully determined by impurities, while the electron-electron and the electron-phonon interactions are responsible for the thermalization of the system.

Fast impurity scattering tends to make the distribution function isotropic, hence it is natural to present the solution of the kinetic equation \eqref{A4} as a sum of an isotropic and anisotropic terms,
\be
\label{A12}
f_\alpha =f_\alpha^i(\ep) +f_\alpha^a(\ep,\hat{\bb{p}}),
\e
where the isotropic part of the distribution function, $f_\alpha^i$, depends only on energy, while the anisotropic part $f_\alpha^a$ depends in addition on the direction of the momentum.

In the impurity-dominated regime the anisotropic part of the collision integral is given by
\be 
\label{St-anis}
\lt({\rm St}[f_\alpha]\rt)^a = -\frac{f_\alpha^a}{\tau(\ep)}, 
\e
where $\tau(\ep)\simeq \tau_{\textrm{imp}}(\ep) $ is the transport scattering time for the electrons (holes) at the energy $\ep$.

To simplify the calculation we consider first an electron-hole symmetric spectrum
\be
\label{A3}
\ep_{e}({\bb{p})}=\ep_{h}({\bb{p})}=\ep_{\bb{p}},
\e
thus assuming that electrons and holes only differ by the sign of charge. A more general situation will be discussed in the next Section. We consider a system with the energy gap $\Delta$ between electron and hole branch of the spectrum, hence $\Delta/2 < \ep_{\bb{p}} < \infty$.

Thus, at charge neutrality, the equilibrium chemical potential equals zero and the corresponding distribution function is given by 
$f_{\bb{p}}^{F}=[1+e^{\ep_{\bb{p}}/T}]^{-1}$ for both electrons and holes.

\textbf{1)} \underline{\textit{Parabolic spectrum and} $\tau = const$.} Let us consider the simplest situation when both electrons and holes have parabolic spectrum: $\ep_{\bb{p}}=\Delta/2+ p^2/2m$, while the impurity scattering rates for both electrons and holes are energy independent and equal: 
$\tau_h(\ep)=\tau_e(\ep)=\tau=const$.

Integrating Eq.~\eqref{A4} over the two-dimensional momentum $\bb{p}$ we arrive at the continuity equations
\be
\label{cont-eh}
\dv{\bb{j}}_e = -\frac{\delta n_h+\delta n_e}{2\tau_R}, \qquad \dv{\bb{j}}_h= -\frac{\delta n_e+ \delta n_h}{2\tau_R},
\e
where we introduce the recombination time $\tau_R$ as well as the non-equilibrium concentrations $\delta n_\alpha$ and the corresponding current densities $\bb{j}_\alpha$,
\be
\label{n,j}
\delta n_\alpha=-\frac{\rho_0}{2}+\int \frac{d^2\bb{p}}{(2\pi)^2}\, f_\alpha , \qquad 
\bb{j}_\alpha= \int \frac{d^2\bb{p}}{(2\pi)^2}\,\bb{v} f_\alpha.
\e
In Eq.~(\ref{n,j}) we took advantage of the definition of the equilibrium quasiparticle density
\be
\rho_0= 2\int\frac{d^2\bb{p}}{(2\pi)^2}\, f^F_{\bb{p}},
\e
which gives the total concentration of all carries (electrons and holes) in equilibrium. 

The term $({\delta n_h+\delta n_e})/{2\tau_R}$ is obtained by the linearization of the phenomenological recombination rate $\gamma n_e n_h$, hence for the symmetric spectrum: $\tau_R^{-1}= 2\gamma n_e^0=2\gamma n_h^0$.

Multiplying Eq.~\eqref{A4} by the velocity $\bb{v}$ and integrating over the momentum we obtain the equation
\be
\label{A16}
\bb{\nabla}\lt[\int \frac{d^2\bb{p}}{(2\pi)^2}\,\frac{v^2}{2}f_\alpha^i\rt] 
-\frac{e_\alpha \bb{E} \rho_0}{2m}-\bb{j}_\alpha \times \bb{\omega}_\alpha=-\frac{\bb{j}_\alpha}{\tau},
\e
where ${\bb{\omega}}_h=-{\bb{\omega}}_e={\bb{\omega}_c}$, $\bb{\omega}_c = e \bb{B}/m$.

In the linear transport regime (linear response) the isotropic part of the distribution function is only slightly deviating from $f^F$. These deviations are of two types: the deviation of the local electronic temperature $\delta T (\bb{r})$ from the lattice temperature and the deviation of the chemical potential $\delta \mu_{\alpha}(\bb{r})$ for electrons and holes from its equilibrium value, which is zero at charge neutrality. 

The temperature of the system is determined by the balance between the Joule heating, recombination, and the cooling by the phonon bath. Since Joule heating is proportional to the square of electric field it can be neglected in linear response. Recombination of electrons and holes would not affect the local temperature as far as the electron-phonon scattering is sufficiently strong:
\be 
\label{A-rec}
\tau_{\textrm{ph}} \ll \tau_R. 
\e
In what follows we assume the inequality \eqref{A-rec} and let $\delta T=0$. Under these assumptions the isotropic part of the distribution function takes the form
\be
\label{delta-f}
f_\alpha^i=f^F-\frac{\pa f^F}{\pa \ep} \delta \mu_\alpha(\bb{r}).
\e
Integrating Eq.~\eqref{delta-f} over the momentum we find
\be
\label{nabla-n}
\bb{\nabla}\delta n_\alpha=\lt\la 1 \rt\ra \bb{\nabla} \delta \mu_\alpha,
\e
where the operation $\la \cdots \ra$ is defined by 
\be
\label{average}
\la \cdots \ra = -\int_{\Delta/2}^{\infty} \!\!\nu(\ep)\,d\ep\;(\cdots)\; \frac{\pa f^F}{\pa \ep},
\e
and $\nu(\ep)$ is the density of states ($\nu(\ep)=const$ for parabolic spectrum).

With the help of Eq.~\eqref{delta-f} one can readily express the integral which enters Eq.~\eqref{A16} in the form
\be
\bb{\nabla} \lt[ \int \frac{d^2\bb{p}}{(2\pi)^2}\,\frac{v^2}{2} f_\alpha^i\rt]=\lt\la \frac{v^2}{2} \rt\ra\,\bb{\nabla}\delta \mu.
\e
Using Eq.~\eqref{nabla-n} we rewrite Eq.~\eqref{A16} as
\be
\label{currents-eh}
D \bb{\nabla} \delta n_\alpha - \frac{e_\alpha E \rho_0\tau}{2m}-\bb{j}_\alpha\times \bb{\omega}_\alpha \tau =-\bb{j}_\alpha.
\e
where we introduce the diffusion coefficient
\be
D=\frac{\la v^2/2\ra\,\tau}{\la 1 \ra},
\e
which is evaluated at charge neutrality as 
\be
\label{D}
\lt.D\rt|_{\mu=0}=\frac{T\tau}{m} \lt(1+e^{\Delta/2T}\rt)\ln\lt(1+e^{-\Delta/2T}\rt).
\e
Introducing the notations $\delta \rho=\delta n_e+\delta n_h$, $\delta n=\delta n_e-\delta n_h$, $\bb{P}=\bb{j}_e +\bb{j}_h$, $\bb{j}= \bb{j}_e -\bb{j}_h$ we readily rewrite Eqs.~(\ref{cont-eh},\ref{currents-eh}) in the form of Eqs.~(7) of the main text. As has been shown in the main text, these equations describe the linear-in-$B$ magnetoresistance at charge neutrality in the limit of strong magnetic field.

\textbf{2)} \underline{\textit{Arbitrary spectrum and} $\tau=\tau(\ep)$.}
The approach developed above is easily generalized for arbitrary spectrum $\ep(\bb{p})$ and arbitrary energy dependence of the scattering time $\tau(\ep)$ (for simplicity, we still consider $\tau$ to be the same for electrons and holes). In this more general case the cyclotron frequency acquires an energy dependence:
\be
\label{A14}
\bb{\omega}_h=-\bb{\omega}_e= \bb{\omega}_c,\qquad  \bb{\omega_c}(\ep)=e\bb{B}\,v/p,
\e
where the dependence of the velocity $v$ and momentum $p$ on energy is implicitly given by the relations  
\be
v(\ep)=\lt|\frac{\pa \ep(p)}{\pa \bb{p}}\rt|,\quad p=|\bb{p}|=p(\ep),\quad \ep_{\bb{p}}=\ep.
\e
Substituting Eq.~\eqref{A12} into Eq.~\eqref{A4} and taking anisotropic part of the obtained equation, we find
\be
\label{kin1}
\bb{v} \bb{\nabla} f_\alpha^i + e_\alpha \bb{E}\bb{v} \frac{\pa f_\alpha^i}{\pa \ep}+\omega_\alpha(\ep) \frac{\pa f_\alpha^a}{\pa \varphi} =
- \frac{f_\alpha^a}{\tau(\ep)},
\e
where $\varphi=\hat{\bb{v}}$ is the velocity angle. 

Integration of Eq.~\eqref{kin1} over momentum would no longer allow us to obtain a closed set of equations on currents $\bb{j}_\alpha$ like it was the case for parabolic spectrum and energy-independent $\tau$. Instead, one has to apply Eq.~\eqref{kin1} in order to express anisotropic part of the distribution function $f_\alpha^a$ via the isotropic one $f_\alpha^i$. The task is accomplished by the relation
\be
f_\alpha^a = \s_{j,k} v^j \tau^{jk}_\alpha \lt(-\frac{\pa}{\pa x^k} - e_\alpha {E^k} \frac{\pa}{\pa \ep}\rt)f_\alpha^i,
\e
where $j,k=x,y$ and $v^j$ ($E^k$) stands for the vector components of velocity (electric field). The tensor components $\tau^{jk}_\alpha=(\hat{\tau}_\alpha)_{jk}$ are arranged into the matrix 
\be
\label{A13}
\hat{\tau}_\alpha=\frac{\tau(\epsilon)}{1+\omega_c^2(\ep)\tau^2(\ep)}
\bpm
1 & \omega_\alpha(\ep) \tau(\ep)\\
-\omega_\alpha(\ep)\tau(\ep)  & 1
\epm.
\e
Multiplying Eq.~\eqref{A13} by the vector $\bb{v}$ and integrating over the velocity angle, we express the electron and hole current densities at a given energy as
\be
\label{jE}
\bb{j}_\alpha(\ep) =- \hat{D}_\alpha^B (\ep) \lt(\bb{\nabla}+e_\alpha \bb{E} \frac{\pa}{\pa \ep}\rt) f_\alpha^i,
\e
where $\hat{D}_\alpha^B(\ep) = v^2 \hat{\tau}_\alpha/2$. With the help of Eq.~\eqref{delta-f} we rewrite Eq.~\eqref{jE} to obtain
\be
\label{A15}
\bb{j}_\alpha(\ep) = \hat{D}_\alpha^B (\ep) \lt[\bb{\nabla}\delta \mu_\alpha(\bb{r})-e_\alpha \bb{E}\rt] \frac{\pa  f^F}{\pa \ep}.
\e
Integrating this equation over the momentum and taking advantage of Eqs.~\eqref{nabla-n} and \eqref{average} we find
\be
\label{j1}
\bb{j}_\alpha = \hat{D}_\alpha^B \lt( - \bb{\nabla} \delta n_\alpha +e_\alpha{\la 1 \ra} {\bb{E}} \rt),
\e
with the tensor
\be
\label{D-ave}
\hat{D}_\alpha^B=\frac{\la \hat{D}_\alpha^B(\ep) \ra}{\la 1 \ra}=
\bpm
D_{\parallel} & \pm D_\perp\\
\mp D_\perp  & D_{\parallel}
\epm,
\e
where the sign $+(-)$ in the upper off-diagonal element stands for holes (electrons), respectively, and 
\beml
\beq
D_{\parallel} &=& \lt\la \frac{v^2}{2} \frac{\tau(\ep)}{1+\omega_c^2(\ep) \tau^2(\ep)}\rt\ra \frac{1}{\la 1\ra},\\
D_\perp &=& \lt\la \frac{v^2}{2} \frac{\omega_c(\ep)\tau^2(\ep)}{1+\omega_c^2(\ep) \tau^2(\ep)}\rt\ra \frac{1}{\la 1\ra}.
\eq
\eml

For energy-independent $\tau$ and $\omega_c$ one finds
\be
\label{D-DB}
\hat{D}_\alpha^B=\frac{D}{1 +\omega_c^2 \tau^2}   
\bpm
1 & \omega_\alpha \tau\\
-\omega_\alpha \tau  & 1
\epm,
\e
where $D$ is given by Eq.~\eqref{D}. Thus, one can indeed restore Eq.~\eqref{currents-eh} directly from Eqs.~\eqref{j1} using the identity $\langle v^2/2\rangle=n_0/m$ for parabolic spectrum. 

In order to find distributions of currents and concentrations for an arbitrary symmetric spectrum we rewrite Eq.~\eqref{j1} in the components by taking advantage that the electron-hole symmetry leads to $\delta n_h=\delta n_e=\delta \rho/2$ and $\bb{E}=E_0\bb{e}_x$ at charge neutrality. Thus, we find
\beml
\label{jxjy}
\beq
\label{jx}
j_{h,x}=-j_{e,x}=e D_\parallel E_0\la 1 \ra - \frac{1}{2}D_\perp\frac{\pa \delta \rho}{\pa y},&&
\\
j_{h,y}=j_{e,y}=-e D_\perp E_0 \la 1 \ra -\frac{1}{2}D_\parallel \frac{\pa \delta\rho}{\pa y}.&& 
\label{jy}
\eq
\eml
Substituting Eqs.~(\ref{jxjy}) into Eqs.~(6) of the main text (since the latter are evidently valid for arbitrary spectrum) we find
\be
\label{ddn}
\frac{\pa^2 \delta \rho}{\pa y^2}  = \frac{4\delta\rho}{\ell_R^2},\qquad  \ell_R = 2\sqrt{D_\parallel/\Gamma}.
\e
Thus, the solution to Eq.~\eqref{ddn} with the boundary condition $j_{\alpha,y}(\pm W/2)=0$ yields
\be
\label{n(y)}
\delta\rho  = -eE_0\ell_R\, \la 1\ra  \frac{D_{\perp}}{D_{\parallel}} \frac{\sinh(2y/\ell_R)}{\cosh(W/\ell_R)}.
\e
Finally, substituting  Eq.~\eqref{n(y)} into Eq.~(\ref{jx}) for $j_{\alpha,x}$ and integrating over $y$ we obtain the total electric current
\be
\label{J}
\overline{J}= 2 e^2 E_0  \lt(D_{\parallel} +\frac{D_{\perp}^2}{D_{\parallel}}\frac{\tanh(W/\ell_R)}{W/\ell_R}\rt).
\e
In the limit  $B \to \infty$ we have
\beq
&&D_\parallel= \frac{1}{(eB)^2} \frac{\lt\la p^2/2\tau\rt\ra}{\la 1 \ra},\quad 
D_{\perp}=\frac{1}{eB}\frac{\lt\la vp/2\rt\ra}{\la 1 \ra},\\
&& \ell_R  = \frac{1}{e B} \sqrt{\frac{2\la p^2/\tau\ra}{\la1\ra \Gamma}}.
\eq
Using these equations we find from Eq.~\eqref{J} the sheet resistance $R_\square = E_0/\overline{J}$,
\be
\lt.R_\square\rt|_{B\to\infty}= \frac{B}{e} \sqrt{\frac{\la 1\ra \la p^2/\tau\ra}{2\tau_R}}\frac{\la 1\ra}{\la vp\ra^2},
\e
which is indeed linear in $B$ for large fields.

\subsection{Beyond charge neutrality and \textit{e-h} symmetry}

Let us now consider the situation when the electron-hole symmetry is absent: i.e. such parameters as mass $m_\alpha$, scattering rate $\tau_\alpha$ and equilibrium concentration $n_{0,\alpha}$ are different for electrons and holes. For simplicity, we will restrict ourselves to a parabolic spectrum, $\ep_\alpha(\bb{p})=\Delta/2+p^2/2 m_\alpha$, assuming that the equilibrium chemical potential is shifted from the middle of the gap, e.g. by doping. In this case, the operation $\la \dots \ra$, which we extensively use to derive the balance equations, is different for electrons and holes and is defined by 
\be
\label{average1}
\langle \cdots \rangle_\alpha =- \int_{\Delta/2}^{\infty} \!\!\nu_\alpha(\ep)d\ep\;  (\cdots) \frac{\pa f_\alpha^F}{\pa \ep},
\e
where $\nu_\alpha(\ep)$ is the corresponding density of states. For parabolic spectrum the diffusion coefficient reads
\be
D_\alpha=\frac{\la v^2/2\ra_\alpha\,\tau_\alpha}{\la 1 \ra_\alpha},
\e
It is clear that away from charge neutrality the diffusion coefficients for electrons and holes become different, even if masses and impurity scattering rates are the same for both types of charge carriers. In this case we readily generalize Eq.~\eqref{D} as 
\be
\label{D1}
D_\alpha=\frac{T\tau}{m} \lt(1+e^{\frac{\Delta/2\pm\mu}{T}}\rt)\ln\lt(1+e^{-\frac{\Delta/2\pm\mu}{T}}\rt),
\e
where we use the convention that the upper-sign in $\pm\mu$ corresponds to the holes ($\alpha=h$), while the lower sign corresponds to the electrons ($\alpha=e$). We regard $\mu$ as the equilibrium chemical potential.

Recombination rates are generally different for electrons and holes and can be approximated as $\Gamma_e= 2\gamma n_h^0$, $\Gamma_h= 2\gamma n_e^0$ assuming the linearisation of the corresponding collision integral. The coefficient $\gamma$ is evidently the function of temperature and depends on a particular model of electron-hole recombination which we do not specify here. 

Repeating the steps outlined in Eqs.~(\ref{cont-eh}-\ref{currents-eh}) for the case of charge neutrality we arrive at Eqs.~(5) of the main text. 
 
In the absence of electron-hole symmetry, the Hall voltage is formed across the sample, which leads in our formulation to the appearance of the $y$-component of the electric field $\bb{E}$, which has to be related to the charge density variation across the sample. In the simple gate approximation we express this relation by Eq.~(6) of the main text. 

The general solution to Eqs.~(5,6) of the main text is not very transparent and is postponed to the next Section of this Supplemental Materials. Before that we discuss two particular cases:  the Boltzmann limit away from charge neutrality and the limit of very fast Maxwell relaxation.

{\bf 1)} \underline{\textit{Boltzmann limit away from charge neutrality}.}
The Boltzmann limit $T \ll \Delta$ corresponds to a small number of charge carriers in both spectrum branches. Let us limit our analysis to the simplest case: $m_\alpha=m,$ $\tau_\alpha=\tau=const$ and, consequently, $\omega_h=-\omega_e=\omega_c$.  Thus, the breaking of electron-hole symmetry is only due to a finite chemical potential $\mu$. In this case the equilibrium distribution functions and concentrations take the form
\beml
\beq
f_e=e^{-\frac{\ep+\Delta/2-\mu}{T}},  &\quad& n_{e,0}= \nu T e^{-\frac{\Delta/2-\mu}{T}},\\
f_h=e^{-\frac{\ep+\Delta/2+\mu}{T}}, &\quad  &n_{h,0}= \nu T e^{-\frac{\Delta/2+\mu}{T}},
\eq
\eml
where $\nu =m/\pi$ is the two-dimensional density of states for a parabolic spectrum. The Drude conductivities and recombination rates for electron and hole subsystems entering Eqs.~(5b) of the main text are given by
\beml
\beq
\sigma_e=\frac{e^2n_e^0 \tau}{m},&\qquad& \Gamma_e=2\gamma n_h^0,\\
\sigma_h=\frac{e^2n_h^0 \tau}{m},&\qquad& \Gamma_h=2\gamma n_e^0.
\eq
\eml    
It follows from Eq.~(\ref{D1}) that the diffusion coefficients in the Boltzmann limit appear to be equal with exponential precision, $D_\alpha =D=T\tau/m$, which strongly simplifies the solution of Eqs.~(5,6) of the main text. Such a solution implies that the $y$-component of the electric field is constant and is given by
\be
\label{Eyy}
E_y= -\frac{\omega_c\tau E_0(\sigma_e-\sigma_h)}{\sigma_h+\sigma_e+DC},
\e
where $C$ is the gate capacitance.  
We also find the sheet resistance $R_\square=E_0/\overline{J}$ as
\be
\label{RBolt}
R_\square=\frac{1}{\sigma_e+\sigma_h}\frac{1+\omega_c^2\tau^2}{1+\omega_c^2\tau^2\lt[\xi +(1-\xi)F(W/\ell_R)\rt]},
\e  
where $\xi=n^2_0/\rho^2_0$ and $\ell_R$ is the magnetic-field dependent recombination length
\be
\ell_R=2\sqrt{\frac{2eD}{(\Gamma_e+\Gamma_h)(1+\omega_c^2\tau^2)}}.
\e    
The result of Eq.~(\ref{RBolt}) is consistent with Eqs.~(11,15) of the main text. 

The appearance of $E_y$ corresponds to a finite Hall voltage $V_\textrm{Hall}=E_yW$. From Eq.~(\ref{Eyy}) we readily calculate the Hall sheet resistance $R_\square^{\textrm{Hall}}=E_y/\overline{J}=R_\square E_y/E_0$ as
\beq
\n
R_\square^{\textrm{Hall}}&=&-\frac{\omega_c\tau}{\sigma_e+\sigma_h+DC}\;\frac{\sigma_e-\sigma_h}{\sigma_e+\sigma_h}\\
&&\times\frac{1+\omega_c^2\tau^2}{1+\omega_c^2\tau^2\lt[\xi +(1-\xi)F(W/\ell_R)\rt]}.
\eq
This completes the analysis of Eqs.~(5,6) of the main text in the Boltzmann limit.

{\bf 2)} \underline{\textit{Fast Maxwell relaxation}.}
Let us turn to a more general situation of different electron and hole masses. The analysis is greatly simplified if we assume that Maxwell relaxation is fast compared to electron diffusion, namely $C\ll m_\alpha e^2$.  In this case one may take the limit $C\to 0$ directly in the Eq.~(6) of the main text, which leads to 
\be 
\label{maxwell}
\delta n_e = \delta n_h =\delta \rho/2.
\e
Thus, we can express the current from Eq.~(5b) of the main text in terms of the concentration and electric field
\be
\label{jjmatrix}
\bb{j}_\alpha = \hat{\Sigma}
\lt(e_\alpha\bb{E} n_{0,\alpha}\tau_\alpha/m_\alpha- D_\alpha \bb{\nabla}\delta n_\alpha\rt).
\e
where
\be
\hat{\Sigma}=\frac{1}{1 +\omega_c^2 \tau_\alpha^2}
\bpm
1 & \omega_\alpha \tau_\alpha \\
-\omega_\alpha\tau_\alpha  & 1
\epm.
\e
To avoid the confusion we remind that  $\omega_\alpha=e_\alpha B/m_\alpha$ have different signs for different $\alpha$. 

The boundary conditions dictate that  $j_{y,e}=j_{y,h}$, therefore one can exclude $E_y$ form Eqs.~(\ref{jjmatrix}) and find the dependence of currents $j_{y,\alpha}$ on $\delta \rho$. Substituting the result into Eqs.~(5,6) of the main text one finds 
\be
\label{drho}
\frac{d^2 \delta\rho}{d^2y}=\frac{4\delta\rho}{\ell_R^2},
\e
where $\ell_R$ is the effective recombination length
\be
\label{kappa1}
\ell_R =2\sqrt{\frac{\sigma_e^{xx} D_h^{xx} + \sigma_h^{xx} D_e^{xx}}{(\Gamma_e+\Gamma_h) (\sigma_e^{xx}+\sigma_h^{xx})}}.
\e
Here we took advantage of the following definitions
\beml
\beq
\hat{\sigma}_\alpha &=& 
\bpm
\sigma_\alpha^{xx} & \sigma_\alpha^{xy}\\
\sigma_\alpha^{yx}  & \sigma_\alpha^{xx}
\epm
=\frac{e^2n_\alpha^0\tau_\alpha}{m_\alpha} \hat{\Sigma},\\
\hat{D}_\alpha(B) &=&
\bpm
D_\alpha^{xx} & D_\alpha^{xy}\\
D_\alpha^{yx}  & D_\alpha^{xx}
\epm
=D_\alpha \hat{\Sigma}.
\eq
\eml
The solution to Eq.~\eqref{drho}, which satisfies the boundary conditions, reads 
\be
\delta\rho=-E_0\ell_R\frac{\sigma_e^{xx}|\sigma_h^{xy}|+|\sigma_e^{xy}|\sigma_h^{xx}}{\sigma_e^{xx}D_h^{xx}+\sigma_h^{xx}D_e^{xx}}\frac{\sinh(2y/\ell_R)}
{\cosh(W/\ell_R)}.
\label{drho1}
\e
Substituting Eq.~\eqref{drho1} into Eq.~\eqref{jjmatrix} we find the inverse sheet resistance
\be 
\label{f1}
R^{-1}_\square=\lt(\rho_{\infty}^{xx}\rt)^{-1}+A F(\kappa W/2), 
\e
where $\rho_{\infty}^{xx}$ is the resistivity of an infinitely large system and 
\be 
\label{A}
A=\frac{(\sigma_e^{xx}|\sigma_h^{xy}|+|\sigma_e^{xy}|\sigma_h^{xx})^2}{(\sigma_e^{xx}+\sigma_h^{xx})\sigma_e^{xx}\sigma_h^{xx}}.
\e
The resistivity tensor in the absence of boundaries is simply given by
\be 
\label{f2}
\hat \rho_\infty =
\bpm
\rho_\infty^{xx} & \rho_\infty^{xy}\\
\rho_\infty^{yx}  & \rho_\infty^{xx}
\epm
=\lt(\hat \sigma_e+\hat \sigma_h\rt)^{-1}.
\e
In the limit of large $B$ we simply have
\be
R_\square=\frac{m}{e^2 \rho_0 \tau}\frac{1}{\ell_R/W  + n_0^2/\rho_0^2}.
\e
Since the recombination length is inversely proportional to magnetic field $\ell_R \propto 1/B$ for large $B$, one can again conclude that $R_\square$ growth linearly with $B$ and saturates when $W/\ell_R$ becomes comparable with $\rho_0^2/n_0^2$ as predicted by Eq.~(15) of the main text. The saturation is obviously absent at charge neutrality $n_0=0$.

\section{General solution}
\label{sec:full}
\subsection{to Eqs.~(5,6) of the main text}

In this Section we obtain the Hall and longitudinal sheet resistance for a rectangular two-dimensional samples with closed boundary conditions at $y=\pm W/2$ by solving Eqs.~(5,6) of the main text. In order to simplify intermediate formulas we replace $\delta n \to n$ and $\delta \rho \to \rho$. We also introduce the following notations
\beml
\beq
\omega_{\pm}=\frac{\omega_e\tau_e\pm\omega_h\tau_h}{2},\qquad\;\;
D_\pm= \frac{D_e\pm D_h}{2},&&\\
\sigma_\pm=\frac{e n_{0,e}\tau_e}{m_e}\pm \frac{en_{0,h}\tau_h}{m_h},\quad
\gamma_{\pm}=\frac{\Gamma_e\pm\Gamma_h}{4}.&&
\eq
\eml
We express Eqs.~(5,6) in terms of currents $\bb{j}=\bb{j}_e-\bb{j}_h$, $\bb{P}=\bb{j}_e+\bb{j}_h$, and densities $n=n_e-n_h$, $\rho=n_e+n_h$ by adding and subtracting the equations for electrons and holes. We note that the continuity equation $\dv \bb{j}=0$ together with the boundary conditions leads to the vanishing $y$ component of the current $\bb{j}$. The other components acquire some $y$ dependence. Thus, we shell use
\be
\bb{j}=(j(y),0),\quad \bb{P}=(P_x(y),P_y(y)). 
\e
Expressing Eq.~(5a) of the main text in the components we obtain
\beml
\label{jPx1}
\beq
\label{jnice}
j &=&\sigma_+E_0+\omega_+P_y,\\
P_x&=&\sigma_-E_0+\omega_-P_y,
\eq
\eml
and
\beml
\label{nrho2}
\beq
(D_+\!+\!\kappa \sigma_+)\frac{\pa n}{\pa y}+D_-\frac{\pa\rho}{\pa y}+\omega_+P_x+\omega_-j=0,\hspace*{36pt}&&\\
(D_-\!+\!\kappa \sigma_-)\frac{\pa n}{\pa y}+D_+\frac{\pa\rho}{\pa y}+P_y+\omega_-P_x+\omega_+j=0,\hspace*{12pt}&&
\eq
\eml
where we introduced $\kappa=e/C$ and took advantage of the Eq.~(6) of the main text in order to exclude the $y$-component of the electric field. We now substitute $j$ and $P_x$ from Eqs.~(\ref{jPx1}) into Eqs.~(\ref{nrho2}) with the result
\beml
\label{nice}
\beq
\label{nicedn}
D_0^2\pa n/\pa y+s_0E_0+b_0P_y &=&0,\\
D_0^2\pa \rho/\pa y+s_1E_0+b_1P_y &=&0,
\eq
\eml
where we introduced more notations
\beml
\beq
D_0&=&\sqrt{D_+(D_++\kappa\sigma_+)-D_-(D_-+\kappa\sigma_-)},\\
s_0&=&(\sigma_+\omega_-+\sigma_-\omega_+)D_+ - (\sigma_+\omega_++\sigma_-\omega_-)D_-,\hspace*{28pt}\\
s_1&=&(\sigma_+\omega_++\sigma_-\omega_-)(D_++\kappa\sigma_+)\n\\
&&-(\sigma_+\omega_-+\sigma_-\omega_+)(D_-+\kappa\sigma_-),\\
b_0&=& 2\omega_+\omega_-D_+ -(1+\omega_+^2+\omega_-^2)D_-,\\
b_1&=&(1+\omega_+^2+\omega_-^2)(D_++\kappa \sigma_+)\n\\
&&-2\omega_+\omega_-(D_-+\kappa\sigma_-).
\eq
\eml
From Eq.~(5b) we also obtain the equation on $\dv\bb{P}$ which we rewrite in the following form
\be
\label{keyE}
\rho=-\frac{1}{\gamma_+}\frac{\pa P_y}{\pa y}-\frac{\gamma_-}{\gamma_+}n.
\e
Now, we substitute $\rho$ from Eq.~(\ref{keyE}) into Eqs.~(\ref{nice}) and use them to exclude $n$. In this way we arrive at the differential equation on $P_y$
\be
\label{diffPy}
\frac{\pa^2 P_y}{\pa^2 y}=\frac{4}{\ell_R^2}P_y+\frac{s_0\gamma_-+s_1\gamma_+}{D_0^2} E_0,
\e
where 
\be
\ell_R =\frac{2D_0}{\sqrt{b_0\gamma_-+b_1\gamma_+}}.\\
\e
The differential equation (\ref{diffPy}) with the boundary conditions $P_y(\pm W/2)=0$ is readily solved. The result has to be averaged over $y$ to obtain the total current,
\be
\overline{P_y}=\big(F(W/\ell_R)-1\big)\frac{s_0\gamma_-+s_1\gamma_+}{b_0\gamma_-+b_1\gamma_+}E_0,
\e
where $F(x)=\tanh(x)/x$ and the averaging is defined as
\be
\overline{P_y}\equiv \frac{1}{W} \int_{-W/2}^{W/2}\!\!\! dy\; P_y(y).
\e
Now the desired relation between $\overline{J}=e\overline{j}$ and $E_0$ is readily obtained by averaging Eq.~(\ref{jnice}) over $y$, hence we find the inverse sheet resistance $R_\square^{-1}=\overline{J}/E_0$ as
\be
\label{finRs}
R_\square^{-1}=e\lt[\sigma_++\omega_+\big(F(W/\ell_R)-1\big)\frac{s_0\gamma_-+s_1\gamma_+}{b_0\gamma_-+b_1\gamma_+}\rt].
\e
Similarly, by averaging the $y$-component of the electric field we obtain with the help of Eq.~(\ref{nicedn})
\beq
\overline{E_y}&=&-\kappa\overline{\frac{\pa n}{\pa y}}=\kappa \frac{s_0E_0+b_0\overline{P_y}}{D_0^2}=\eta\,E_0,\\
\eta&=&\frac{\kappa}{D_0^2}\lt[s_0+b_0\big(F(W/\ell_R)-1\big)\frac{s_0\gamma_-+s_1\gamma_+}{b_0\gamma_-+b_1\gamma_+}\rt],\hspace*{16pt}
\eq
Thus, the Hall sheet resistance is simply found as
\be
\label{finRH}
R_\square^{\textrm{Hall}}=\overline{E_y}/\overline{J} = \eta\,R_\square.
\e
The results of Eqs.~(\ref{finRs},\ref{finRH}) are plotted in Fig.~3 of the main text using realistic parameters.

\section{Interaction-dominated regime}
\subsection{Hydrodynamic approach}

In this Section we discuss the hydrodynamic regime assuming the following hierarchy of the scattering rates
\be
\label{A1}
\tau_{ee}^{-1} \gg \tau^{-1}_{imp},~ \tau_{ph}^{-1},
\e
where $\tau_{ee}^{-1}$ is the characteristic rate for the electron-electron collisions, while $\tau^{-1}_{imp} $ and $\tau_{ph}^{-1}$ are the impurity and the electron-phonon transport scattering rates, respectively. For simplicity, we restrict ourselves to an electron-hole symmetric spectrum, $\ep_{\alpha}(\bb{p})=\ep_{\bb{p}}$ at charge neutrality.

One may think that inequality of Eq.~\eqref{A1} ensures that both liquid components are fully characterised by local spatially-dependent temperature $T(\bb{r})$,  chemical potentials $\mu_\alpha(\bb{r})$, and drift velocities $\bb{u}_\alpha(\bb{r})$.  This would imply the following Ansatz for the distribution function 
\be
\label{A2}
f_{\alpha}=\frac{1}{\exp\lt[\lt(\ep_{\bb{p}}- \bb{p} \bb{u}_{\alpha}(\bb{r}) - \mu_{\alpha}(\bb{r})\rt)/ {T(\bb{r})}\rt] +1},
\e
which nullifies the electron-electron part of the collision integral. In that logic the equations for hydrodynamic functions $\bb{u}_{\alpha}(\bb{r})$, $\mu_{\alpha}(\bb{r})$,  and $T(\bb{r})$ are found by substituting Eq.~\eqref{A2} into Eq.~\eqref{A4}. The closed system of equations is, then, obtained by multiplying Eq.~\eqref{A4} by $1$, $\bb{p}$, and $\ep_{\bb{p}}$ with the subsequent integration over the momentum. 

Unfortunately, such a program is not legitimate for two-liquid systems with a realistic collision integral, since the latter almost always implies nearly equal collision rates for electron-electron, hole-hole, and electron-hole scattering. This suggests a notable friction between electron and hole components of the liquid when electric field is applied. The friction force is proportional to the difference in drift velocities: $\bb{u}_{e}-\bb{u}_{h}$. Since electrons and holes move in opposite direction the friction makes the velocities vanish in the hydrodynamic limit.  In order to obtain a nonzero current one has to go beyond the hydrodynamic approximation and study the corrections to Eqs.~\eqref{A2}. The rigorous approach to the problem for arbitrary spectrum leads to cumbersome equations \cite{Brooker1972}. The analysis somewhat simplifies for materials with linear spectrum such as graphene due to the enhanced forward scattering \cite{Kashuba2008,Fritz2008,Schuett2011}. Still, even in the case of linear spectrum, the derivation of a closed set of hydrodynamic equations lacks physical transparency. 

In order to simplify the analysis we focus here on the model situation assuming that the rate of the electron-hole scattering is low compared to the rates of electron-electron and hole-hole scattering: $\tau_{eh}^{-1} \ll \tau_{hh}^{-1}=\tau_{ee}^{-1}$. In this case, one can still use the Ansatz of Eq.~\eqref{A2} because of the fast equilibration within each liquid component. For simplicity we again limit ourselves to the case of a parabolic spectrum assuming that the impurity scattering time is energy independent. Generalisation of the theory for arbitrary spectrum is, then, straightforward.

In linear response it is legitimate to expand the distribution function as 
\be
\label{A5}
f^\alpha=f^F+\delta f^\alpha, 
\e
where
\be
\label{A6}
\delta f^\alpha = -\frac{\pa f^F}{\pa \ep}  \lt(\delta \mu_\alpha +  \ep_{\bb{p}}\,\delta T/T  + \bb{p}\, \bb{u}_\alpha \rt),
\e
and $\delta \mu$, $\delta T$, and $\bb{u}$ are proportional to the electric field $\bb{E}$. As far as the cooling rate associated with the phonons ($\propto 1/\tau_{ph}$) is faster than recombination rate $1/\tau_R$ we may disregard the temperature fluctuations, $\delta T=0$. In this limit  the concentration of electrons and holes are related to the variation of chemical potential $\delta \mu_\alpha$ by means of Eq.~\eqref{nabla-n}, while the currents are proportional to hydrodynamic velocities
\be
\bb{j}_\alpha= m\langle v^2\rangle \bb{u}_\alpha/2=\la \ep-\Delta/2 \ra \bb{u}_\alpha.
\e
Integrating Eq.~\eqref{A4} over the momentum we obtain the continuity equations \eqref{cont-eh}. Integration with velocities yields, however, the equations
\beml
\label{currents-eh1}
\beq
\label{jh}
D \bb{\nabla} \delta n_h  -\frac{e E_0 \rho_0\tau}{2m}- \bb{j}_h\times \bb{\omega_c} \tau -\bb{F}_{eh} =-{\bb{j}_h},&&  \\
\label{je}
D \bb{\nabla} \delta n_e  +\frac{e E_0 \rho_0\tau}{2m}+ \bb{j}_e\times \bb{\omega_c} \tau +\bb{F}_{eh} =-{\bb{j}_e},&&
\eq
\eml
which differ from Eq.~\eqref{currents-eh} only by the presence of the friction force
\be
\bb{F}_{eh}=\chi (\bb{j}_e-\bb{j}_h)/2,
\e
where $\chi \simeq \tau/\tau_{eh}$. It is worth noting that the hydrodynamic approach implies $\tau/\tau_{\textrm{e-e}} \gg 1$ and $\tau/\tau_{hh} \gg 1$, while the parameter $\chi$ can take on arbitrary values as far as $\tau_{eh}\gg \textrm{max}\{\tau_{ee},\tau_{hh}\}$.

The electron-hole symmetry dictates the following relations at charge neutrality: $\delta n_h=\delta n_e$, $j_{x,e}=-j_{x,h}=j/2$, and $j_{y,e}=j_{y,h}=P/2$.
Using these relations we transform Eqs.~(\ref{cont-eh},\ref{jh},\ref{je}) into the following set of equations 
\beml
\label{system}
\beq
e E \rho_0\tau/m-(1+\chi)j+\omega_c\tau P&=&0,\\
2D\;\pa\delta n/\pa y+P+\omega_c\tau j&=&0,\\
\pa P/\pa y&=&\delta n/\tau_R,
\eq
\eml
which is supplemented by the boundary conditions $P(\pm W/2)=0$.  From the solution of Eqs.~(\ref{system}) we get
\beml
\beq
\label{n-beta}
n&=&-\frac{eE_0\ell_R\rho_0\tau}{4m} \frac{\omega_c\tau}{D(1+\chi)} \frac{\sinh(2y/\ell_R)}{\cosh(W/\ell_R)},\\
\overline{J}&=&\frac{e^2E_0\rho_0\tau}{m(1+\chi)}\frac{1+ \chi+ \omega_c^2\tau^2 F(W/\ell_R)}{1+\chi+\omega_c^2\tau^2},
\eq
\eml
where
\be
\ell_R=2\sqrt{\frac{(1+\chi) D\tau_R}{1+\chi +\omega_c^2\tau^2}}.
\e
In the limit of large magnetic field $\omega_c\tau \gg \sqrt{1+\chi}$ and for $W \gg \ell_R$  we again obtain linear-in-B magnetoresistance
\be
R_\square= E_0/\overline{J}=\frac{\sqrt{1+\chi}}{2e\rho_0\sqrt{D\tau_R}}\,B.
\e
In conclusion we note that Eqs.~(5) of the main text can be generalised by including the friction force $\bb{F}_{eh}$. The most general solution of these equations would, then, anyway lead to a linear non-saturating magnetoresistance at charge neutrality.


\begin{thebibliography}{99}

\bibitem{Friedman2010}
A.\,L.~Friedman, J.\,L.~Tedesco, P.\,M.~Campbell, J.\,C.~Culbertson, E.\,Aifer, F.\,K.~Perkins, R.\,L.~Myers-Ward, J.\,K.~Hite, C.\,R.~Eddy, G.\,G.~Jernigan, and D.\,K.~Gaskill, Nano Lett. \textbf{10}, 3962 (2010).

\bibitem{Singh2012} 
R.\,S.~Singh, X.\,Wang, W.\,C.~Ariando, and A.\,T.\,S.~Wee, App.\ Phys.\ Lett.  \textbf{101}, 183105 (2012).

\bibitem{Veldhorst2013} 
M.\,Veldhorst, M.\,Snelder, M.\,Hoek, C.\,G.~Molenaar, D.\,P.~Leusink, A.\,A.~Golubov, H.\,Hilgenkamp, and A.\,Brinkman,
Phys.\ Status Solidi RRL \textbf{7}, 26 (2013).

\bibitem{Wang2013} 
W.\,Wang, Y.\,Du, G.\,Xu, X.\,Zhang, E.\,Liu, Z.\,Liu, Y.\,Shi, J.\,Chen, G.\,Wu, and X.\,Zhang, Sci.\ Rep. \textbf{3}, 2181 (2013).

\bibitem{Gusev2013}
G.\,M.~Gusev, E.\,B~Olshanetsky, Z.\,D.~Kvon, N.\,N.~Mikhailov, and S.\,A.~Dvoretsky, Phys.\ Rev.\ B \textbf{87}, 081311(R) (2013).

\bibitem{Wiedmann2013}
S.\,Wiedmann, Private Communications (2013).

\bibitem{Weiss1954} 
H.\,Weiss and H.\,Welker, Z.\ Phys. \textbf{138}, 322 (1954).

\bibitem{Kapitza1928} 
P.\,L.~Kapitza, P.\ L.\ Proc.\ R.\ Soc.\ London A \textbf{119}, 358 (1928).

\bibitem{Yang1999}
F.\,Y.~Yang, K.\,Liu, K.\,Hong, D.\,H.~Reich, P.\,C.~Searson, and C.\,L.~Chien, Science \textbf{284}, 1335 (1999).

\bibitem{Yang2000}
F.\,Yang, K.\,Liu, K.\,Hong, D.\,Reich, P.\,Searson, C.\,Chien, Y.\,Leprince-Wang, K.\,Yu-Zhang, and K.\,Han, Phys.\ Rev.\ B \textbf{61}, 6631 (2000).

\bibitem{Xu1997}
R.\,Xu, A.\,Husmann, T.\,F.~Rosenbaum, M.-L.\,Saboungi, J.\,E.~Enderby, and P.\,B.~Littlewood, Nature \textbf{57}, 390 (1997).

\bibitem{Husmann2002} 
A.\,Husmann, J.\,B.~Betts, G.\,S.~Boebinger, A.\,Migliori, T.\,F.~Rosenbaum, and M.-L.\,Saboungi, Nature \textbf{417}, 421 (2002).

\bibitem{Sun2003} 
Y.\,Sun, M.\,B.~Salamon, M.\,Lee, and T.\,F.~Rosenbaum, Appl.\ Phys.\ Lett. \textbf{82}, 1440 (2003).

\bibitem{Zhang2004}
X.\,Zhang, Q.\,Z.~Xue, and D.\,D.~Zhu, Phys.\ Lett. A \textbf{320}, 471 (2004).

\bibitem{Hu2008}
J.\,Hu and T.\,F.~Rosenbaum, Nat.\ Mat. \textbf{7}, 698 (2008).

\bibitem{Dalven1967}
R.\,Dalven and R.\,Gill, Phys.\ Rev.\ B \textbf{159}, 645 (1967).

\bibitem{Liang2014}
T.\,Liang, Q.\,Gibson, M.\,N.~Ali, M.\,Liu, R.\,J.~Cava, and N.\,P.~Ong, arXiv:1404.7794 (2014).

\bibitem{Ali2014}
M.\,N.~Ali, J.\,Xiong, S.\,Flynn, Q.\,Gibson, L.\,Schoop, N.\,Haldolaarachchige, N.\,P.~Ong, J.\,Tao, and R.\,J.~Cava, Nature \textbf{514}, 205 (2014).

\bibitem{Wang2014}
K.\,Wang, D.\,Graf, and C.\,Petrovic, arXiv:1405.1719 (2014).

\bibitem{Pletikosic2014} 
I.\,Pletikosic, M.\,N.~Ali, A.\,Fedorov, R.\,J.~Cava, and T.\,Valla, arXiv:1407.3576 (2014).

\bibitem{Abrikosov1969}
A.\,A.~Abrikosov, Sov.\ Phys.\ JETP \textbf{29}, 746 (1969); Phys.\ Rev.\ B \textbf{58}, 2788 (1998); Europhys.\ Lett. \textbf{49}, 789 (2000).

\bibitem{Parish2003}
M.\,M.~Parish and P.\,B.~Littlewood, Nature \textbf{426}, 162 (2003).

\bibitem{Kittel1963}%
C.\,Kittel, \textit{Quantum Theory of Solids} (Wiley, New York, 1963).

\bibitem{Titov2013}
M.\,Titov, R.\,V.~Gorbachev, B.\,N.~Narozhny, T.\,Tudorovskiy, M.\,Schuett, P.\,M.~Ostrovsky, I.\,V.~Gornyi, A.\,D.~Mirlin, M.\,I.~Katsnelson, K.\,S.~Novoselov, A.\,K.~Geim, L.\,A.~Ponomarenko, Phys.\ Rev.\ Lett. \textbf{111}, 166601 (2013).

\bibitem{Chang2014}
S.-J.\,Chang, M.\,Bawedin, and S.\,Cristoloveanu,
IEEE Trans.\ Electr.\ Dev. \textbf{61}, 1979 (2014).

\bibitem{sup} Supplementary Material

\bibitem{long}
M. Sch\"utt, M. Titov, I.V. Gornyi, B.N. Narozhny, and A.D. Mirlin,
to be published.

\bibitem{Knap2014}
M.\,Knap, J.\,D.~Sau, B.\,I.~Halperin, and E.\,Demler,
arXiv:1405.0277 (2014).

\end{thebibliography}

\begin{thebibliography}{99}

\bibitem{Brooker1972}
G.\,A.~Brooker and J.\,Sykes, Annals of Physics, \textbf{74}, 67 (1972).

\bibitem{Kashuba2008}
A.\,B.~Kashuba, Phys.\ Rev.\ B \textbf{78}, 085415 (2008).

\bibitem{Fritz2008} 
L.\,Fritz,  J.\,Schmalian, M.\,M\"{u}ller, and S.\,Sachdev, Phys.\ Rev.\ B \textbf{78}, 085416 (2008).

\bibitem{Schuett2011}
M.\,Sch\"{u}tt, P.\,M.~Ostrovsky, I.\,V.~Gornyi, and A.\,D.~Mirlin, Phys.\ Rev.\ B \textbf{83}, 155441 (2011).

\end{thebibliography}
\end{document}